\documentclass[11pt,sort&compress]{elsarticle}

\journal{}
\usepackage{lipsum}
\usepackage{amsfonts}
\usepackage{graphicx}
\usepackage{epstopdf}
\usepackage{algorithmic}
\usepackage{multirow}
\usepackage{subcaption}
\usepackage{epsf}
\usepackage{amsmath}
\usepackage{amssymb}
\usepackage{setspace}
\usepackage{enumerate}
\usepackage{float}
\usepackage{footnote}
\usepackage{scalefnt}
\usepackage{xfrac}
\usepackage{transparent}
\usepackage{rotate}
\usepackage{array}
\usepackage{tabu}
\usepackage{multirow}
\usepackage{enumitem}
\usepackage{tikz}
\usepackage{bm}
\usepackage{microtype}
\usepackage[margin=1in]{geometry}
\usepackage{listings}
\usepackage{color}
\usepackage{url}

\usepackage{siunitx}
\DeclareSIUnit{\atmosphere}{atm}
\newcommand\blfootnote[1]{%
  \begingroup
  \renewcommand\thefootnote{}\footnote{#1}%
  \addtocounter{footnote}{-1}%
  \endgroup
}

\usepackage{bm}
\newcommand{\ve}[1]{\bm{#1}}

\newcommand{\bu}{\ve{u}}

\newcommand{\bq}{\ve{q}}

\newcommand{\bh}{\ve{h}}
\newcommand{\bs}{\ve{s}}

\newcommand{\bF}{\ve{F}}
\newcommand{\bI}{\ve{I}}

\usepackage{xcolor}
\definecolor{lightblue}{rgb}{0.63, 0.74, 0.78}
\definecolor{seagreen}{rgb}{0.18, 0.42, 0.41}
\definecolor{orange}{rgb}{0.85, 0.55, 0.13}
\definecolor{silver}{rgb}{0.69, 0.67, 0.66}
\definecolor{rust}{rgb}{0.72, 0.26, 0.06}

\colorlet{lightsilver}{silver!30!white}
\colorlet{darkorange}{orange!75!black}
\colorlet{darksilver}{silver!65!black}
\colorlet{darklightblue}{lightblue!65!black}
\colorlet{darkrust}{rust!85!black}

\definecolor{dkgreen}{rgb}{0,0.6,0}
\definecolor{gray}{rgb}{0.5,0.5,0.5}
\definecolor{mauve}{rgb}{0.58,0,0.82}

\usepackage{hyperref}
\hypersetup{
  colorlinks=true,
}

\usepackage{cleveref}
\crefname{lstlisting}{listing}{listings}
\Crefname{lstlisting}{Listing}{Listings}

\definecolor{dkgreen}{rgb}{0,0.6,0}
\definecolor{gray}{rgb}{0.5,0.5,0.5}
\definecolor{mauve}{rgb}{0.58,0,0.82}

\usepackage[labelfont=bf,justification=raggedright,singlelinecheck=false]{caption}
\usepackage{parskip}

\begin{document}

\hypersetup{
  linkcolor=darkrust,
  citecolor=darklightblue,
  urlcolor=darkrust,
}

\begin{frontmatter}

\title{%
{\large\bfseries Method for scalable and performant GPU-accelerated \\ simulation of multiphase compressible flow}}

\author[1]{Anand~Radhakrishnan}
\author[1]{Henry Le~Berre}
\author[1]{Benjamin Wilfong}
\author[2]{Jean-Sebastien~Spratt}
\author[3]{Mauro~Rodriguez~Jr.}
\author[2]{Tim~Colonius}
\author[1,4]{Spencer H.\ Bryngelson\corref{cor1}}
\ead{shb@gatech.edu}
\cortext[cor1]{Corresponding author}

\address[1]{School of Computational Science \& Engineering, 
Georgia Institute of Technology, Atlanta, GA 30332, USA}
\address[4]{Daniel Guggenheim School of Aerospace Engineering, Georgia Institute of Technology, Atlanta, GA 30332, USA}

\address[2]{Division of Engineering and Applied Science, California Institute of Technology, Pasadena, CA 91125, USA}

\address[3]{School of Engineering, Brown University, Providence, RI 02912, USA}

\date{}

\begin{abstract}
    Multiphase compressible flows are often characterized by a broad range of space and time scales, entailing large grids and small time steps.
    Simulations of these flows on CPU-based clusters can thus take several wall-clock days.
    Offloading the compute kernels to GPUs appears attractive but is memory-bound for many finite-volume and -difference methods, damping speedups.
    Even when realized, GPU-based kernels lead to more intrusive communication and I/O times owing to lower computation costs.    
    We present a strategy for GPU acceleration of multiphase compressible flow solvers that addresses these challenges and obtains large speedups at scale. 
    We use OpenACC for directive-based offloading of all compute kernels while maintaining low-level control when needed.
    An established Fortran preprocessor and metaprogramming tool, Fypp, enables otherwise hidden compile-time optimizations. 
    This strategy exposes compile-time optimizations and high memory reuse while retaining readable, maintainable, and compact code.
    Remote direct memory access realized via CUDA-aware MPI and GPUDirect reduces halo-exchange communication time.
    We implement this approach in the open-source solver MFC~\citep{bryngelsonMFCOpensourceHighorder2021}.
    Metaprogramming results in an 8-times speedup of the most expensive kernels compared to a statically compiled program, reaching $46\%$ of peak FLOPs on modern NVIDIA GPUs and high arithmetic intensity (about 10~FLOPs/byte).
    In representative simulations, a single NVIDIA A100 GPU is $7$-times faster compared to an Intel Xeon Cascade Lake (6248) CPU die, or about $300$-times faster compared to a single such CPU core.
    At the same time, near-ideal ($97$\%) weak scaling is observed for at least 13824 GPUs on OLCF Summit. 
    A strong scaling efficiency of $84$\% is retained for an 8-times increase in GPU count. 
    Collective I/O, implemented via MPI3, helps ensure the negligible contribution of data transfers ($<1\%$ of the wall time for a typical, large simulation). 
    Large many-GPU simulations of compressible (solid-)liquid-gas flows demonstrate the practical utility of this strategy.
\end{abstract}


\end{frontmatter}

\section{Introduction}

\blfootnote{All code available at \url{https://github.com/MFlowCode/MFC}}

Multiphase compressible flows are ubiquitous, with examples such as the atomization of liquid droplets~\citep{mengNumericalSimulationsDroplet2016}, bubble cavitation~\citep{brennen2015cavitation}, and shock-wave attenuation of nuclear blasts~\citep{chauvin2011experimental}. 
Collapsing bubble clouds in cavitating flows can result in shock waves that lead to large pressures. 
This has applications in a wide variety of engineering problems, such as the design of mechanical heart valves~\citep{johansen2004mechanical}, burst-wave lithotripsy~\citep{katzInvestigationEnergyShielding2018}, and minimizing blast-induced trauma~\citep{movahedCavitationinducedDamageSoft2016}. 
These high pressures also cause erosion of industrial equipment in flow around hydrofoils~\citep{seoNumericalInvestigationCloud2009}, pumps~\citep{dagostinoCavitationInstabilitiesRotordynamic2017}, and propellers~\citep{jofreTranscriticalDiffuseinterfaceHydrodynamics2021}. 
Simulation of multiphase phenomena is thus critical to enable engineering design and minimize equipment damage.

Courant--Friedrichs--Lewy (CFL) constraints for compressible flow restrict permissible time step sizes.
Thus, many time steps are needed to simulate the relevant physical phenomena, making the minimization of wall time for each time step critical. 
Since 2004, CPU clock speeds have plateaued, ending Dennard scaling.
Thus, the compute capabilities of modern supercomputers stem primarily from GPU accelerators.
Leveraging GPUs is thus essential to extracting meaningful speedups from state-of-the-art supercomputers. 

Multiphase compressible flow algorithms consist mostly of level 1 BLAS (vector) stencil operations, so most are memory bound~\citep{datta2009optimization}. 
The low arithmetic intensity of these kernels prevents efficient use of the GPU's compute capabilities. 
Faster GPU kernels can also lead to prominent MPI communication and I/O transfer times. 
This can exacerbate strong scaling behavior for smaller problem sizes with large numbers of GPUs. 
We present strategies that address the concerns above to obtain satisfactory performance on accelerators. 
The portability of our acceleration strategies is ensured by conducting tests on various architectures. 
Large multi-GPU simulations are included to emphasize the benefits of our strategies in pertinent applications. 

GPU speedups, scaling tests, and example simulations are conducted with the open-source solver MFC~\citep{bryngelsonMFCOpensourceHighorder2021}. 
Interface capturing methods~\citep{saurelDiffuseInterfaceCapturingMethods2018}, particularly the $5$- and $6$- equation models~\citep{kapila2001, allaireFiveEquationModelSimulation2002, richardsaurelSimpleEfficientRelaxation2009}, are used to represent the multi-component flow.
These equations are discretized and solved using a shock-capturing finite volume scheme that uses high-order accurate WENO reconstructions~\citep{coralicFinitevolumeWENOScheme2014, jiangEfficientImplementationWeighted1996}. 
The Riemann problem is then solved using an HLLC approximate Riemann solver~\citep{toroRiemannSolversNumerical2009}, and the solution is evolved with a total-variation-diminishing (TVD) Runge--Kutta time stepper~\citep{gottliebTotalVariationDiminishing1998}.  
The GPU acceleration strategies outlined in this work take advantage of the increased arithmetic intensity of high-order accurate methods typically used for multicomponent flow simulations.  

Large-scale compressible flow simulations have been conducted on CPUs for a long time and still gather substantial attention~\citep{rasthofer2019computational}.
However, attempts at optimizing such solvers for GPU acceleration are less unified as the GPU hardware landscape evolves.
We cite solvers like STREAmS (version~1 and~2)~\citep{bernardini2021streams,bernardini2023streams} and ZEFR~\citep{romero2020zefr} as just a couple of demonstrative examples of these efforts.
STREAmS simulates compressible turbulent flow via a flux vector splitting method and achieves $250$-times speedups on a single NVIDIA~V100 GPU over an Intel Skylake CPU core. 
They observe $97\%$ weak scaling efficiency for up to 1024~V100 GPUs along with $90\%$ strong scaling efficiency for an 8-times increase in GPU count. 
ZEFR employs similar strategies and observes similar accelerations, simulating single-phase compressible flows and retaining $70\%$ strong scaling efficiency for an 8-times increase in GPU count.
These solvers do not address the challenges associated with multicomponent flows, which is part of our focus.

The flexibility associated with the GPU programming model is a concern of increasing importance. 
CUDA offers reliable performance on NVIDIA GPUs and has been a GPU-programming mainstay since its inception in 2007.
However, other vendors like AMD and Intel are deploying competitive GPU accelerators in the most capable new supercomputers, like OLCF Frontier, CSC LUMI, and ALCF Aurora~\citep{melesse2022approaching,zwinger2023lumi,jiang2022intel}. 
Vendor-specific programming models like CUDA are insufficient to take advantage of the capability these new supercomputers bring.
Deploying performant and vendor-agnostic fluid flow solvers on new computers requires adopting more flexible programming models.
Here, we use OpenACC~\citep{wienke2012openacc}, a performance-competitive directive-based model with established support for NVIDIA GPUs~\citep{khalilov2021performance} and increasing support for AMD and Intel hardware~\citep{jarmusch2022analysis}.
The FluTAS solver~\citep{crialesi2023flutas} also adopted OpenACC, solving the incompressible multiphase flow problem via a finite difference scheme. 
However, speedups are limited in the incompressible flow case due to communication times associated with pressure-Poisson solves and Fourier transforms.
In their study, FluTAS displayed linear weak scaling and a $40\%$ retention in strong scaling efficiency for an 8-times increase in GPU count on the MeluXina supercomputer~\citep{varrette2021uni}.
URANOS~\citep{de2023uranos} uses OpenACC similarly for turbulent compressible flows, demonstrating the approach's efficacy for compressible CFD applications.
However, the algorithms employed degrade performance by $20\%$ when weak scaling up to 300~GPUs, and the speedups are limited compared to the ones we present here.

We describe the computational models used to formulate the governing equations in \cref{s:model}.
The numerical method that solves the discrete conservation laws is outlined in \cref{s:num}. 
\Cref{s:impl} describes optimal GPU acceleration and MPI communication strategies.
Results for model validation, GPU speedups, and scaling tests are presented in \cref{s:main}. 
The benefits of GPU acceleration for large multiphase problems are illustrated via example simulations in \cref{s:exmp}. 
\Cref{s:conclusions} highlights the relevant conclusions from this work. 

\section{Computational model}\label{s:model}

We briefly describe the multicomponent models used in GPU acceleration tests. 
These models are reduced from the non-equilibrium Baer--Nunziato model~\citep{andrianov2004riemann}. 

\subsection{5-equation models}\label{ss:all}

The so-called Kapila 5-equation model~\citep{kapila2001} is obtained from the non-equilibrium Baer--Nunziato model~\citep{andrianov2004riemann} under the assumptions of velocity and pressure equilibrium between the phases. 
The equations for 2 components are
\begin{align*}
    \frac{\partial \alpha_1 \rho_1}{\partial t} + \nabla \cdot (\alpha_1 \rho_1 \bu) &= 0, \\
    \frac{\partial \alpha_2 \rho_2}{\partial t} + \nabla \cdot (\alpha_2 \rho_2 \bu) &= 0, \\
    \frac{\partial \rho \bu}{\partial t} + \nabla \cdot (\rho \bu \otimes \bu + p \bI) &= 0, \\
    \frac{\partial \rho E}{\partial t} + \nabla \cdot [(\rho E + p)\bu] &= 0, \\
    \frac{\partial \alpha_1}{\partial t} + \bu \cdot \nabla \alpha_1 &= K \nabla \cdot \bu,
\end{align*}
where $\rho$, $\bu$, $p$, and $E$ are the mixture density, velocity, pressure, and energy, and $\alpha_i$ are the volume fractions of component $i$.
The system of equations is closed using an equation of state (EOS).
Here, we use the stiffened gas EOS, which can faithfully model many liquids and gases~\citep{le2016noble}:
\begin{gather} \label{eqn:soe}
    \rho E = \frac{1}{\gamma - 1} p + \frac{\gamma \pi_{\infty}}{\gamma - 1},
\end{gather}
though other relations can be used as appropriate.
For a 2-component problem, we have
\begin{gather}
    K = \frac{\rho_2 c_2^2 - \rho_1 c_1^2}{\frac{\rho_2 c_2^2}{\alpha_2} + \frac{\rho_1 c_2^2}{\alpha_1}},
\end{gather}
and $K \nabla \cdot \bu$ represents the expansion and compression of each phase in mixture regions and ensures thermodynamic consistency via the conservation of phase entropy.
This admits a consistent representation of the sound speed in the mixture region, though it can lead to numerical instabilities due to the non-conservative source term in the volume fraction advection equation~\citep{coralicFinitevolumeWENOScheme2014}.

The $K\nabla \cdot\bu$ term can be ignored in cases where mixture compression effects are unimportant, though it is unclear how to determine this a priori.
For example, a case where they are known to be important is spherical bubble dynamics~\citep{schmidmayerAssessmentMulticomponentFlow2020}.
If one can ignore this term, the equations degenerate to the Allaire model~\citep{allaireFiveEquationModelSimulation2002}.
Though the Allaire model is conservative, it does not strictly obey the second law of thermodynamics. 

\subsection{A 5-equation model with hypoelasticity}\label{ss:hyp}

A hypoelastic material model represents the elastic response of solids~\citep{rodriguez_2019}.
The model is obtained by modifying the 5-equation model.
An elastic shear stress term $\tau_{ij}^{(e)}$ modifies the Cauchy stress tensor as
\begin{gather}
    \sigma_{ij} = -p\delta_{ij}+\tau^{(v)}_{ij} + \tau_{ij}^{(e)},
\end{gather}
where $\boldsymbol{\tau}^{(v)}$ is the viscous stresses.
An elastic contribution $e^{(e)}$ contributes to the total energy $E$ as
\begin{gather}
    E = e + \frac{\|\bu\|^2}{2} + e^{(e)} \quad \text{where} \quad e^{(e)} = \frac{(\tau_{ij}^{(e)})^2}{4 \rho G}.
\end{gather}
Additional equations are required to track the elastic stresses.
In 3D, this is 6 additional equations, one for each stress term $\tau_{ij}^{(e)}$ where $i,j \in \{1,2,3\}$ and $\boldsymbol{\tau}^{(e)}$ symmetric.
With elastic contributions and additional equations, the hypoelastic 5-equation model for 2 materials is

\begin{align*}
    \frac{\partial \alpha_1 \rho_1}{\partial t} + \nabla \cdot (\alpha_1 \rho_1 \bu) &= 0, \\
    \frac{\partial \alpha_2 \rho_2}{\partial t} + \nabla \cdot (\alpha_2 \rho_2 \bu) &= 0, \\
    \frac{\partial \rho \bu}{\partial t} + \nabla \cdot (\rho \bu \otimes \bu + p \bI)  + \nabla \cdot (\boldsymbol{\tau}^{(e)} +  \boldsymbol{\tau}^{(v)}) &= 0, \\
    \frac{\partial \rho E}{\partial t} + \nabla \cdot [(\rho E + p)\bu] - \nabla \cdot [(\boldsymbol{\tau}^{(e)} +  \boldsymbol{\tau}^{(v)}) \bu]  &= 0, \\
    \frac{\partial \alpha_1}{\partial t} + \bu \cdot \nabla \alpha_1 &= K \nabla \cdot \bu, \\
    \frac{\partial \tau^{(e)}_{il}}{\partial t} + \nabla \cdot (\tau^{(e)}_{il} \bu) &= S^{(e)}_{il},
\end{align*}
where 
\begin{gather}
    S^e_{il} = 
    \rho \left(\tau_{kj}^{(e)}\frac{\partial u_i}{\partial x_k} + 
    \tau_{ik}^{(e)}\frac{\partial u_j}{\partial x_k} - 
    \tau_{ij}^{(e)}\frac{\partial u_k}{\partial x_k} + 
    2G\dot{\epsilon}_{ij}^{(d)}\right).
\end{gather}

\subsection{6-equation model with $p$-relaxation}\label{ss:sau}

The numerical instabilities introduced by the Kapila model can be alleviated via the pressure disequilibrium model of \citet{richardsaurelSimpleEfficientRelaxation2009}. 
The system of equations is first evolved without the source terms, followed by a pressure relaxation step under the assumption of infinite stiffness for the pressure relaxation coefficient as discussed in \citet{schmidmayerAssessmentMulticomponentFlow2020}. 

\section{Numerical method} \label{s:num}

Herein, we describe the numerical method that evaluates the governing $5$/$6$-equation models of \cref{s:model}. 

\subsection{Finite volume method (FVM)} \label{ss:fvm}

A finite volume numerical scheme that follows~\citet{coralicFinitevolumeWENOScheme2014} is used to discretize the governing equations
\begin{gather} 
    \frac{\partial \bq}{\partial t} + \frac{\partial \bF^x(\bq)}{\partial x} + \frac{\partial \bF^y(\bq)}{\partial y} + \frac{\partial \bF^{z}(\bq)}{\partial z} = \bs(\bq) - \bh(\bq) \nabla \cdot \bu,
    \label{eqn:fvm}
\end{gather}
where $\bq$ and $\bF$ represent the conservative variables and fluxes in the governing equations.     
The finite volume method represents the conservative variables $\bq_{i,j,k}$ centered at the location $(x_i, y_j, z_k)$. 
The dimensions of the cell are
\begin{gather}
    I_{i,j,k} = [x_{i-1/2}, x_{i+1/2}] \times   [y_{j-1/2}, y_{j+1/2}] \times  [z_{k-1/2}, z_{k+1/2}],
\end{gather}
with grid spacing 
\begin{gather}
    \Delta x_i = x_{i+1/2} - x_{i-1/2}, \quad \Delta y_j = y_{j+1/2} - y_{j-1/2}, \quad \Delta z_k = z_{k+1/2} - z_{k-1/2}. 
\end{gather}

The PDE \eqref{eqn:fvm} is integrated in space across each cell center as
\begin{equation} \label{eqn:fvmdis}
\begin{aligned}
    \frac{\partial \bq_{i,j,k}}{\partial t} = \frac{1}{\Delta x_i}(\bF^x_{i-1/2,j,k} - \bF^x_{i+1/2,j,k}) + \frac{1}{\Delta y_j}(\bF^y_{i,j - 1/2,k} -  \bF^y_{i,j + 1/2,k}) \\ + \frac{1}{\Delta z_k}(\bF^z_{i,j,k - 1/2} - \bF^z_{i,j,k + 1/2}) + \bs(\bq_{i,j,k}) - \bh(\bq_{i,j,k})(\nabla \cdot \bu)_{i,j,k},
\end{aligned}
\end{equation}
where $\bq_{i,j,k}$ are the volume averaged conservative variables, and $\bs(\bq_{i,j,k})$ and $\bh(\bq_{i,j,k})$ are the volume averaged source terms.
The flux term $\bF^x (\bq_{i+1/2,j,k})$ is obtained by averaging over the finite volume cell face $A_{i+1/2,j,k}$ centered at $\left( x_{i+1/2}, y_j, z_k \right)$. 
The other flux terms in \eqref{eqn:fvmdis} are obtained using a similar procedure. 
The divergence term is computed as 
\begin{equation}
    \begin{aligned}
    (\nabla \cdot \bu)_{i,j,k} = \frac{1}{\Delta x_i} (u^x_{i+1/2,j,k} - u^x_{i-1/2,j,k}) + \frac{1}{\Delta y_j} (u^y_{i,j+1/2,k} - u^y_{i,j-1/2,k}) \\ + \frac{1}{\Delta z_k} (u^z_{i,j,k+1/2} - u^z_{i,j,k-1/2}),
    \end{aligned}
\end{equation}
where $u^x_{i+1/2,j,k}$ is the $x$ direction of the velocity averaged across the cell face at grid point $A_{i+1/2, j, k}$.
     
\subsection{Shock capturing via reconstruction}\label{ss:weno}

The numerical scheme in~\cref{ss:fvm} requires the reconstruction of the fluxes $\bF$ and the velocity $\bu$ at the cell faces. 
The state variable $\bq$ at a cell face $A_{i+1/2,j,k}$ is reconstructed from the face's left and right sides, resulting in a discontinuity.
The resulting flux and velocity at the cell face are obtained by solving a Riemann problem at the interface~\citep{menikoff1989riemann}:  
\begin{align*}
    \bF^x_{i+1/2,j,k} &= \bF^x(\bq^\mathrm{L}_{i+1/2,j,k}, \bq^\mathrm{R}_{i+1/2,j,k}) \\ 
    \bu^x_{i+1/2,j,k} &= \bu^x(\bq^\mathrm{L}_{i+1/2,j,k}, \bq^\mathrm{R}_{i+1/2,j,k}).
\end{align*}
The superscripts L and R denote the reconstructed state variable at the left and right cell faces. 
A first-order, total variation diminishing approximation of the state variables at the interface $A_{i+1/2,j,k}$ follows as
\begin{gather}
    \bq^\mathrm{L}_{i+1/2,j,k} = \bq_{i,j,k} \quad \text{and} \quad  \bq^\mathrm{R}_{i+1/2,j,k} = \bq_{i+1,j,k}.
\end{gather}

This scheme suppresses spurious oscillations at interfaces, but can lose accuracy via their smearing.
For this, high-order accurate reconstructions at the interface help keep interfaces sharp for a given grid resolution.
Here, we will show results for a fifth-order-accurate WENO reconstruction, though the method is also performant for the third-order accurate variant.

WENO reconstruction follows from a convex combination of interpolating polynomials on candidate stencils. 
A $(2k-1)$th-order WENO reconstructed state variable $f_{i+1/2,j,k}$ is obtained by a weighted sum of $k$ candidate polynomials as
\begin{gather}
    f_{i+1/2,j,k} = \sum_{r=0}^k \omega^r_{i+1/2} f^r_{i+1/2,j,k}.
\end{gather}
The weights $\omega^r$ are obtained from ideal weights $w^r$ using smoothness indicators $\beta^r$. 
Additional implementation details for WENO3 and WENO5 are in \citet{shuNumericalMethodsHyperbolic2006}.   
High-order WENO reconstructions are not total variation diminishing (TVD).
This makes it susceptible to spurious oscillations at material interfaces. 
To suppress these oscillations, the conservative variables $\bq$ are converted to the primitive ones before reconstruction~\citep{coralicFinitevolumeWENOScheme2014}. 

\subsubsection{Approximate Riemann solver}

The Riemann problem in~\cref{ss:weno} is solved using an approximate HLLC (Harten-Lax-van Leer contact) Riemann solver~\citep{toro2019hllc}. 
The HLLC Riemann solver admits three discontinuities in the solution with wave speeds $s_\mathrm{L}$, $s_*$, and $s_\mathrm{R}$. 
The resulting state at the cell interface $A_{i+1/2,j,k}$ is given as
\begin{equation}
    \bq_{i+1/2,j,k} =
    \begin{cases}
        \bq^\mathrm{L}_{i+1/2,j,k} & 0 \leq s_\mathrm{L} \\
        \bq^{\mathrm{L}*}_{i+1/2,j,k} & s_\mathrm{L} \leq 0 \leq s_* \\
        \bq^{\mathrm{R}*}_{i+1/2,j,k} & s_* \leq 0 \leq s_\mathrm{R} \\
        \bq^\mathrm{R}_{i+1/2,j,k} & 0 \geq s_\mathrm{R}
    \end{cases}.
\end{equation}
The wave speeds $s_\mathrm{L}$ and $s_\mathrm{R}$ are estimated using the state variables $\bq^\mathrm{L}_{i+1/2,j,k}$ and $\bq^\mathrm{R}_{i+1/2,j,k}$. 
In order to calculate the intermediate states $\bq^{\mathrm{L}*}_{i+1/2,j,k}$ and $\bq^{\mathrm{R}*}_{i+1/2,j,k}$, continuity of normal velocity ($u^{\mathrm{L}*} = u^{\mathrm{R}*} = u^*$) and pressure ($p^{\mathrm{L}*} = p^{\mathrm{R}*} = p^*$) is imposed across the contact discontinuity with speed $s_* = u^*$.
The flux at the cell interface is then calculated as
\begin{equation}
    \bF(\bq_{i+1/2,j,k}) =
    \begin{cases}
        \bF(\bq^\mathrm{L}_{i+1/2,j,k}) & 0 \leq s_\mathrm{L} \\
        \bF(\bq^\mathrm{L}_{i+1/2,j,k}) + s_\mathrm{L}(\bq^{\mathrm{L}*}_{i+1/2,j,k} - \bq^\mathrm{L}_{i+1/2,j,k}) & s_\mathrm{L} \leq 0 \leq s_* \\
        \bF(\bq^\mathrm{R}_{i+1/2,j,k}) + s_\mathrm{R}(\bq^{\mathrm{R}*}_{i+1/2,j,k} - \bq^\mathrm{R}_{i+1/2,j,k}) & s_* \leq 0 \leq s_\mathrm{R} \\
        \bF(\bq^\mathrm{R}_{i+1/2,j,k}) & 0 \geq s_\mathrm{R} 
    \end{cases}.
\end{equation}

\subsection{Boundary conditions}

Boundary conditions are implemented by allocating buffers at the domain edges.
Buffer sizes are determined based on the order of WENO reconstruction.
Boundary conditions for time-dependent hyperbolic systems require knowledge of solutions exterior to the computational domain. 
We use characteristic-based boundary conditions for this purpose \citep{thompsonTimedependentBoundaryConditions1990}.

\subsection{Time stepping}

Once the right-hand side of~\eqref{eqn:fvmdis} is determined using spatial reconstruction coupled with a Riemann solver, the solution is evolved in time by discretizing the time derivative.
For this purpose, a high-order total variation diminishing Runge--Kutta time stepper~\citep{gottliebTotalVariationDiminishing1998} achieves temporal accuracy while suppressing spurious oscillations. 
    
\section{Implementation strategy} \label{s:impl}

\subsection{Domain decomposition and I/O}
    
Distributed computing techniques are essential to improve performance at large scales. 
This is achieved by decomposing the domain into 3D blocks across multiple processors. 
The block dimensions for each processor are kept uniform across all dimensions instead of splitting across a single dimension (slabs). 
This is done to minimize the data communicated at the processor boundaries.     

We utilize a structured mesh with non-uniform spacing to discretize the domain. 
Cartesian and Cylindrical geometries are supported, with the axisymmetric option available for cylindrical grids.
For 3D cylindrical geometries, fast Fourier transforms maintain finite spacing along the circumferential direction near the center. 
This is implemented with the FFTW package on CPUs and cuFFT on NVIDIA GPUs~\citep{frigo1998fftw}. 
Grid stretching through a hyperbolic tangent function is done to locally refine the grids near locations of interest~\citep{vinokur1983one}. 

High-order reconstruction at processor boundaries requires knowledge of the state variables outside the local domain.  
Communication of the boundary data among adjacent processors, called the halo exchange, is thus required across all dimensions at each time step. 
Blocking send and receive MPI calls transfer data in the halo region~\citep{gropp1999using}. 
Data in the halo regions are packed into 1D buffers for compatibility with the MPI subroutines and then unpacked into the requisite data structures. 
Ideal weak and strong scaling is observed on multiple CPU cores~\citep{bryngelsonMFCOpensourceHighorder2021}. 

Unsteady compressible flow simulations require I/O data saves at a fixed number of time steps to obtain the temporal variation of the solution.
We use a parallel I/O framework for the file systems, which enables collective MPI read and write functions~\citep{thakur1999data}. 
I/O data dumps can thus be performed in a distributed manner across all processors,
ensuring scalability of the code on modern supercomputers. 
Silo files are exported from the data for analysis and visualization~\citep{collette2010silo}. 

\subsection{GPU offloading}

All computation within a time step is offloaded to GPUs via OpenACC~\citep{wienke2012openacc}.
After applying the initial condition in the pre-processing step, the state variables are copied to the GPUs. 
The state variables are transferred back to the CPUs only during I/O data saves, typically occurring once every one thousand steps for large problems. 
Directive-based offloading in OpenACC only requires the specification of the location of independent loops and the level of parallelization. 
This allows the compiler to decide the optimum kernel parameters for maximum speedup. 
Another advantage of directive-based offloading is maintaining a common codebase for the CPU and GPU versions and enabling accelerators via a compiler flag. 
cuTENSOR and cuFFT libraries are used for optimized tensor reshapes and Fourier transforms on GPUs.

\begin{lstlisting}[caption={A Fortran90+Fypp code snippet of the WENO5 reconstruction kernel.},label={weno},xleftmargin = 0mm,float] 
#:for NORM_DIR, dir in [(1, 'x'), (2, 'y'), (3, 'z')]
if (norm_dir == ${NORM_DIR}$) then
    !$acc parallel loop gang vector collapse (3) private(dvd, poly, beta, alpha, omega)
    do l = is3%beg, is3%end !third coordinate direction
        do k = is2%beg, is2%end !second coordinate direction
            do j = is1%beg, is1%end !first coordinate direction
                !$acc loop seq
                do i = 1, sys_size
                    !$acc loop seq
                    do q = -2, 1 !compute divided differences
                        dvd(q) = v_${dir}$(j + q + 1, k, l, i) &
                                - v_${dir}$(j + q, k, l, i)
                    end do
                    !$acc loop seq
                    do q = 0, 2
                        poly(q) = v_${dir}$(j, k, l, i) &
                                 + poly_coef_${dir}$(j, q, 0)*dvd(q+1) &
                                 + poly_coef_${dir}$(j, q, 1)*dvd(q  )
                        beta(q) = beta_coef_${dir}$(j, q, 0)*dvd(1-q)*dvd(1-q) &
                                 + beta_coef_${dir}$(j, q, 1)*dvd(1-q)*dvd( -q) &
                                 + beta_coef_${dir}$(j, q, 2)*dvd( -q)*dvd( -q) 
                    end do                    
                    alpha = d_${dir}$(:, j)/(beta**2)
                    omega = alpha/sum(alpha)                    
                    vL_${dir}$(j, k, l, i) = sum(omega*poly)                    
                end do
            end do
        end do
    end do
end if
#:endfor
\end{lstlisting}

Our OpenACC implementation uses well-established directives, making it portable.
The code is compatible with NVHPC, GNU, and Cray (CCE) compilers on NVIDIA GPUs, with incoming support for AMD GPUs for GNU and CCE.
Here, we conduct simulations on NVIDIA V100 (OLCF Summit) and A100 GPUs (OLCF Wombat) using the NVHPC compiler.
CPU simulations are also supported on various architectures using NVHPC and GCC compilers with appropriate optimization flags. 
We test CPU simulations on Intel Xeon Cascade Lake CPUs (PSC's Bridges2), IBM POWER9 CPUs (OLCF Summit), and ARM CPUs (OLCF Wombat). 

A code snippet example of an OpenACC kernel used in MFC is given in~\cref{weno}. 
The kernel uses 5th-order-accurate WENO to reconstruct the state variables at the cell faces. 
WENO reconstruction constitutes about $40\%$ of the total time step on GPUs, making it the most expensive OpenACC kernel. 
Various optimization techniques used in~\cref{weno} to improve kernel performance are detailed in~\cref{ss:opt}. 

High-order WENO reconstruction of state variables entails larger stencils.
Larger stencils are commensurate with larger halo regions at the processor boundaries and longer communication times. 
Halo exchange times on GPUs can be significant owing to faster kernel times over CPUs. 
Moreover, the proportion of data on the processor boundaries increases with decreasing problem size. 
Minimizing MPI communication time is thus essential to retain GPU speedups for smaller problem sizes.
For this purpose, we use CUDA-aware MPI and GPUDirect RDMA to accelerate communication by using fast GPU interconnects~\citep{wang2013gpu}.
We observe 4-times performance improvement for halo exchanges using CUDA-aware MPI over conventional MPI communication with Spectrum MPI $10.4$ on OLCF Summit. 

\subsection{Optimization} \label{ss:opt}

OpenACC kernels use gangs, workers, and vectors to distribute the workload, which maps to blocks, warps, and vectors in CUDA notation. 
The \texttt{parallel loop} constructs in OpenACC split the loop iterations across gangs by default.
This usually leads to each block using a single OpenACC vector, resulting in a huge waste of resources.
The \texttt{parallel loop} construct in~\cref{weno} is thus augmented with a \texttt{gang vector} clause that enables the splitting of the loop iterations across multiple gangs of fixed vector length~\citep{chandrasekaran2017openacc}. 
The compiler chooses the optimum number of gangs and vectors for efficient resource allocation~\citep{chandrasekaran2017openacc}. 
Furthermore, the three loops across multiple dimensions in~\cref{weno} are combined into a single loop using the \texttt{collapse(3)} clause.  
This allows the compiler to choose optimal gang and vector sizes for a specific architecture based on the total problem size, thus eliminating the effect of any skewness in the grid dimensions. 
The fourth loop in~\cref{weno} benefits from serialization due to its smaller loop bounds, with $\texttt{sys\_size}$ typically between 5 and 9 for multiphase problems. 

Compressible flow algorithms are primarily vector operations, typically leading to low arithmetic intensity. 
Potential for improved GPU performance was observed in the computationally intensive WENO reconstruction kernel, outlined in~\cref{weno}, through a high degree of memory reuse. 
Flattened multidimensional arrays are preferred over user-defined data types, allowing for aggressive compiler optimizations in GPU kernels. 
A $6$-times performance improvement was observed using multidimensional arrays for the kernel in~\cref{weno} over user-defined data types for a 3D problem with 1M ($10^6$) grid points. 

Memory coalescence is needed to saturate the GPU's global memory bandwidth. 
This can be achieved by ensuring that successive threads in a parallelized loop access consecutive memory locations. 
Previous serialized implementations benefited from switching the inner and outer loops in~\cref{weno} corresponding to iteration variables \texttt{j} and \texttt{k}. 
This allowed for reusability of the coefficient variables (\texttt{poly\_coef}, \texttt{beta\_coef}) across the iterations of the inner loop. 
However, reordering the two loops, as in~\cref{weno}, ensured thread coalescence across inner iterations and proved beneficial for GPU implementation. 

Dimensional splitting in the finite volume method requires independent reconstructions and flux additions across all dimensions. 
This allows for temporarily reshaping the state variables to achieve coalesced memory access. 
The additional cost of reshaping state variables is mitigated by the high degree of reuse of the reshaped variables in the computationally intensive GPU kernels. 
The state variables (\texttt{v}, \texttt{vL}) in \cref{weno} are reshaped so that the fastest-changing index \texttt{j} matches the direction of reconstruction.  
Reshaping the state variables for the kernel in~\cref{weno} results in a $10$-times speedup for a 3D problem with 1M grid points per NVIDIA V100 GPU. 
The cuTENSOR library reshapes these arrays on NVIDIA GPUs, though we observe similar performance when reshaping manually via array loops.

Metaprogramming techniques, in this case, enabled by the Fypp pre-processor~\citep{fypp}, 
 further improves GPU kernel performance. 
It allows for user inputs to be passed as compile-time constants, which are used to allocate fixed-size thread-local arrays in GPUs.
Their availability as fixed-size arrays allows the compiler to introduce additional optimizations in the kernel and thus primarily (fast) register memory accesses.
Fixed parameters are also used to eliminate large conditional blocks in kernels, leading to improved times via judicious use of available GPU registers.
Kernel duplication across multiple dimensions in \cref{weno} is also eliminated through the use of Fypp macros (\texttt{\#:for~NORM\_DIR}). 
This strategy results in 8- and 2-times speedups of the two most expensive kernels associated with WENO and the Riemann solver, respectively.
OpenACC does not automatically inline serial subroutines within GPU kernels. 
This can cause a $10$-times slowdown of kernels with sequential subroutine calls.  
Fypp enables manual inlining of these subroutines to retain compiler optimizations and speedups. 
Fypp is only a metaprogramming tool, so it does not generate any code that a programmer could not write themselves.
However, it does write code a programmer is \textit{unlikely} to write themselves out of the sheer work and repeated optimizations required.
In the case of MFC, fypp also enables a reduced line count by about a factor of 10.

The same core kernel code is used for CPUs and GPUs.
After optimization, we found a minor speedup in the CPU simulations before GPU porting.
Thus, no meaningful performance difference is observed on CPUs by optimizing for GPUs via OpenACC offloading.

\subsection{Validation}

MFC has been validated on test cases via comparisons to experimental results for various physical problems such as shock-bubble interaction, shock-droplet, spherical bubble collapse, and Taylor--Green vortices~\citep{bryngelsonMFCOpensourceHighorder2021}. 
The results are verified to be high-order accurate by monitoring error convergence. 
Solutions are also free of any spurious oscillations at material interfaces.

\section{Performance results}\label{s:main}

The 5-equation model in \cref{ss:all} is used to obtain performance metrics for both CPUs and GPUs. 
Similar GPU speedups and scaling are observed for the $5$-equation hypoelastic model in \cref{ss:hyp} and the 6-equation model in \cref{ss:sau}.

\subsection{Scaling}

The scalability of the code is probed by examining wall times for a 3D 2-component problem with air and water on OLCF Summit. 
The hardware details of Summit are available in \citet{vergara2019scaling}.
In brief, Summit is an IBM system of AC922 POWER nodes.
It contains 4600 compute nodes, each with 2~IBM~POWER9 processors and 6~NVIDIA~V100 GPUs.
Each POWER9 chip is connected via NVLINK with a bandwidth of $25$~GB/s.
Each node has $96$~GB of HBM2 GPU memory.
The wall times are averaged across $10$ time steps during a simulation, which is sufficient for variability to be sufficiently small that it is visually indistinguishable on subsequent plots.
The processor with maximum wall time is used to examine results.  

\subsubsection{Weak scaling} 

\begin{figure}
    \centering
    {\setlength{\tabcolsep}{-0.3cm}
    \begin{tabular}{c c}
        \includegraphics[scale=1]{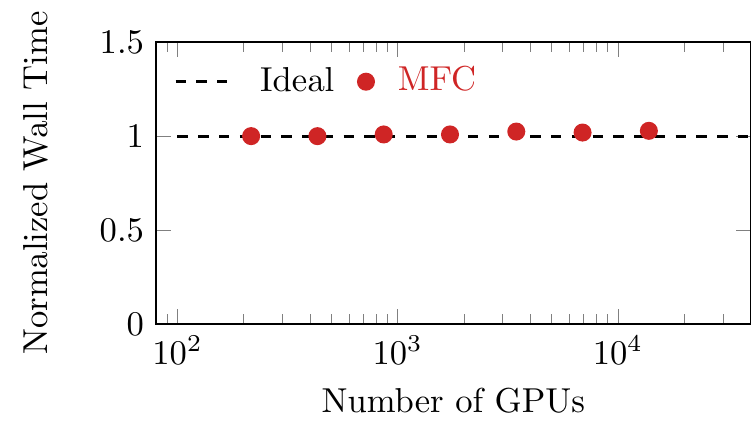} &
        \includegraphics[scale=1]{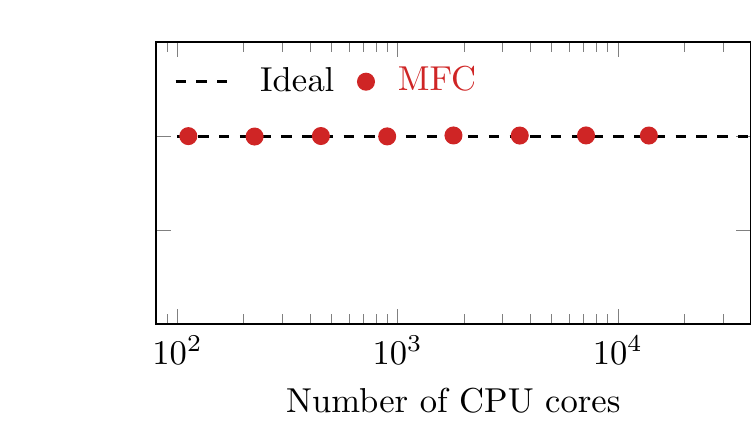} \\
        \small (a) GPU Scaling & \small (b) CPU scaling
    \end{tabular}
    }
    \caption{Weak scaling for a 3D 2-phase problem on (a) NVIDIA V100 GPUs and (b) POWER9 CPUs.}
    \label{fig:weak}
\end{figure}

The problem size per GPU or CPU is 1M grid points in 3D with 1 MPI rank per processor. 
Here, the problem size increases proportionately to the number of processors to test weak scaling.
\Cref{fig:weak} shows the weak scaling performance using NVIDIA V100 GPUs and POWER9 CPUs on OLCF Summit. 
The wall times are normalized using the base case with $216$ GPUs and $112$ CPU cores, respectively. 
A near-ideal efficiency of $97$\% is observed for at least 13824 GPUs. 
Efficiency is within 1\% of ideal performance for at least 14336 CPU cores. 
A higher weak scaling efficiency on CPUs over GPUs can be attributed to the negligible contribution of halo exchanges.

\subsubsection{Strong scaling} 

We use the same 3D 2-component problem as the weak scaling test on Summit to observe strong scaling performance. 
This is achieved by fixing the total problem size and increasing the number of processors. 
The base case uses 8 GPUs or CPU cores, with 1 MPI rank per core.
For GPU cases, this corresponds to one core binding to each GPU.
We use problem sizes of $64$ and $16$M grid points with the base case using $8$ and $2$M grid points per GPU or CPU, respectively. 
For simulations with GPU quantities that are not multiple of 6, the extra GPUs are spilled onto the next node.

\begin{figure}[h]
    \centering
    \includegraphics[scale=1]{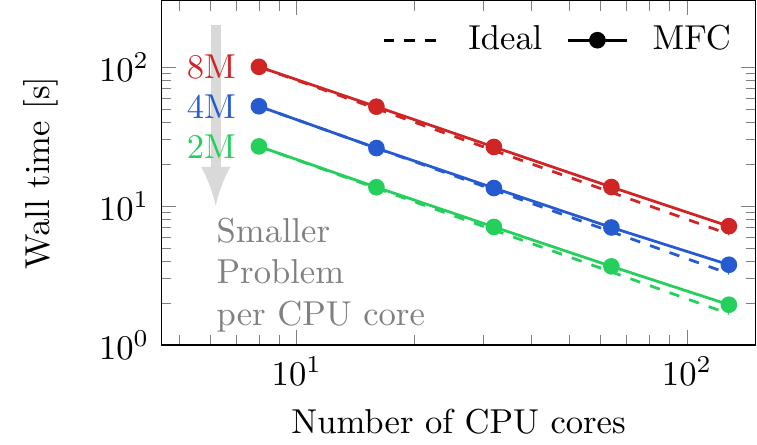} 
    \caption{Strong scaling results on POWER9 CPUs for a 3D 2-component problem with sized as labeled.}
    \label{fig:strong_cpu}
\end{figure}

\begin{figure}[h]
    \centering
    {\setlength{\tabcolsep}{0.4cm}
    \begin{tabular}{ c c }
        \includegraphics[scale=1]{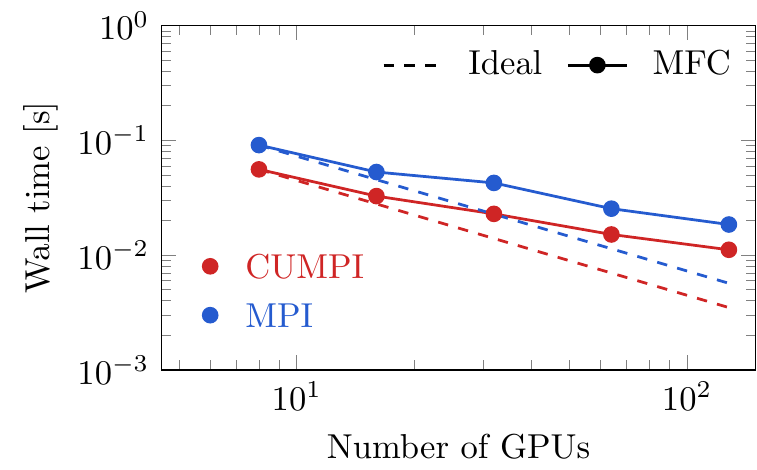} &
        \includegraphics[scale=1]{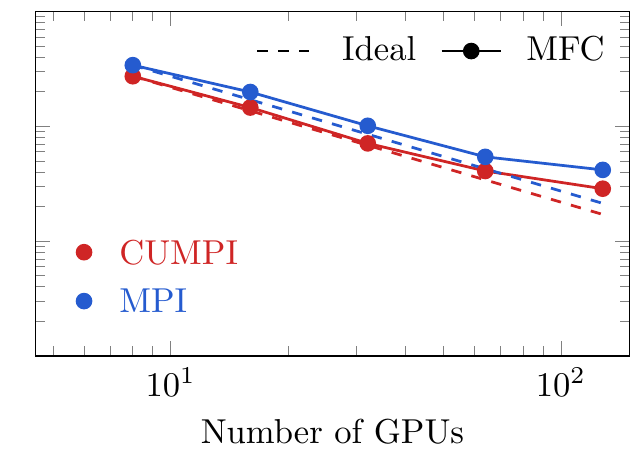} \\
        \small (a) 16M Grid points & \small (b) 64M Grid points
    \end{tabular}
    }
    \caption{Strong scaling analysis on NVIDIA V100 GPUs with and without CUDA-aware MPI (CUMPI) for a 3D 2-component problem with problem sizes (number of grid points) as labeled.}
    \label{fig:strong_gpu}
\end{figure}

\Cref{fig:strong_cpu} shows strong scaling performance on POWER9 CPUs on OLCF Summit.
We observe ideal scaling varying from one to six nodes. 
This indicates that communication is negligible for CPU simulations and problems of appreciable size, which is discussed further in \cref{ss:sys_prof}.
This contrasts against GPU simulations, for which computational time is markedly shorter but network communication times are not.
\Cref{fig:strong_gpu} shows strong scaling performance using NVIDIA V100 GPUs with and without GPU-aware MPI on OLCF Summit. 
An increase in processor count leads to decreased problem size per GPU. 
Deviation from ideal performance is expected due to the increasing share of MPI communication. 

\begin{table}[h]
    \centering
    \renewcommand{\arraystretch}{1.15}
    \small
    \begin{tabular}{ r | r r | r r }
        \hline\hline
          &  \multicolumn{2}{c}{CUMPI}    & \multicolumn{2}{c}{MPI} \\ 
        GPUs  & Ideal [s] & Actual [s] & Ideal [s] & Actual [s] \\ \hline
        8   & 0.273\phantom{0} & 0.273\phantom{0} & 0.341\phantom{0} & 0.341\phantom{0} \\
        16  & 0.146\phantom{0} & 0.137\phantom{0} & 0.199\phantom{0} & 0.171\phantom{0} \\
        32  & 0.0710 & 0.0683 & 0.101\phantom{0} & 0.0853 \\
        64  & 0.0408 & 0.0341 & 0.0542 & 0.0426 \\
        128 & 0.0286 & 0.0171 & 0.0417 & 0.0213 \\
        \hline\hline
    \end{tabular}
    \caption{
        Strong scaling simulation times for the 64M grid point case of \cref{fig:strong_gpu}~(b).
    }
    \label{t:strong_scaling_gpu}
\end{table}

CUDA-aware MPI and the GPUDirect RDMA use fast GPU interconnects to achieve up to 4-times faster halo exchanges. 
This can be observed for the largest problem size (64M) in~\cref{fig:strong_gpu} and in more detail in \cref{t:strong_scaling_gpu}, retaining $84\%$ of ideal performance for a factor of 8 increase in processor count. 
A larger deviation from ideal performance is observed without CUDA-aware MPI.
The problem size is chosen to be close to the memory limit of an NVIDIA V100-16GB SMX2 GPU. 
However, deviations from ideal behavior are observed for smaller problem sizes (16M) at larger numbers of GPUs (64 and 128). 
This can be attributed to MPI communication dominating the simulation time ($>50$\% of wall time) as the compute required per GPU diminishes more quickly but the communication does not.
Of course, one expects a full plateau and nearly 100\% of the time spent doing communication for a sufficiently large number of GPUs.

\subsection{Profiles and I/O} \label{ss:sys_prof}

We monitor the contribution of various subroutines in the code to the total simulation time for GPUs and CPUs.
This is done by measuring the wall times of the most expensive routines for a single time step. 
The wall times are averaged across 10-time steps, and the processor with maximum wall time is used for runs with multiple processors. 
\Cref{tab:rout} shows the percentage contribution of most expensive subroutines for a 3D 2-component problem using 4~NVIDIA V100 GPUs on OLCF Summit. 
CPU simulations use GNU compilers with \texttt{-Ofast} optimization flag. 
We test a relatively small problem with 4M grid points (1M per processor or accelerator) and a larger problem with 32M grid points (8M per processor or accelerator). 
The larger problem size is chosen to be close to the hardware limit of an NVIDIA V100-16GB SMX2 GPU (\SI{16}{\giga\byte}). 

WENO reconstruction is observed to be the most expensive kernel in MFC, taking around $40$\% and $70$\% of the time step for GPUs and CPUs.
A relatively smaller contribution of the WENO kernel on GPUs indicates large speedups for this kernel. 
MPI communication time on GPUs has a decreasing contribution to the overall simulation time with increasing problem size. 
The contribution of MPI communication on CPUs is relatively insignificant ($\approx 1$\%) and invariant to problem size. 
It should be noted that I/O data saves, while insignificant on CPUs, require the equivalent of about $10$ time step units on GPUs. 
However, CFL-restricted time step sizes mean data saves are typically conducted once every 1000 steps.
Hence, simulation times remain largely unaffected by I/O transfers on all architectures. 

\begin{table}
    \centering
    \renewcommand{\arraystretch}{1.15}
    \small
    \begin{tabular}{r c c c c }
        \hline\hline
        Subroutine  & GPU(1M)\% & GPU(8M)\% &  CPU(1M)\% & CPU(8M)\%  \\
        \hline
        Reconstruction  & 37.0 & 43.3 & 72.9 & 71.7 \\
        Riemann Solve   & 24.7 & 33.3  &  16.4 & 18.7 \\
        Communication   & 17.6 & 5.81 & 3.64 & 3.83 \\
        Other           & 20.7 & 17.6 & 7.06 & 5.77 \\
        \hline\hline
    \end{tabular}
    \caption{Percentage contributions of the most expensive subroutines per time step. 
    The GPUs are NVIDIA V100s, and the CPUs are IBM POWER9s. 
    The problem size per CPU or GPU as labeled (1M and 8M).}
    \label{tab:rout}
\end{table}

\subsection{Kernel performance} \label{s:kern}

The analysis of system profiles in~\cref{ss:sys_prof} indicates that WENO and Riemann solver kernels account for over $75$\% of the simulation time on GPUs for large problem sizes ($8$M per GPU). 
Individual speedups of the most expensive subroutines using GPUs over CPUs for the larger simulation in~\cref{ss:sys_prof} are given in~\cref{tab:speedup_Rout}.   
Large speedups ($480$-times) for the most expensive WENO kernel on GPUs result from coalesced memory access and a high degree of reuse. 
The Riemann solvers kernel achieves lower speedups on GPUs due to a large number of thread-local variables and a low degree of memory reuse.  
Speedups in the MPI subroutines are observed due to faster buffer packing and unpacking on GPUs and CUDA-aware MPI.
A $300$-times speedup is observed for the time step using a single V100 GPU over a POWER9 CPU core.
On a single Summit compute node, this translates to a $40$-times speedup using the GPUs over only the CPUs.

\begin{table}
    \centering
    \renewcommand{\arraystretch}{1.15}
    \small
    \begin{tabular}{r l}
        \hline\hline
        Subroutine  & Speedup   \\
        \hline
         Reconstruction & 486  \\
         Riemann Solver & 189 \\
         Communication  & 125 \\
         \hline
         Total & 305\\
         \hline\hline
    \end{tabular}
    \vspace{0.2cm}
    \caption{
        Speedups (times faster) for the most expensive subroutines and the overall time step using V100 GPUs over a POWER9 CPU for a 3D 2-component problem with 8M grid points on OLCF Summit. 
        This problem uses 4~GPUs and 4~CPU cores. 
    }
    \label{tab:speedup_Rout}
\end{table}

\begin{figure}
    \centering
    \includegraphics[scale=1]{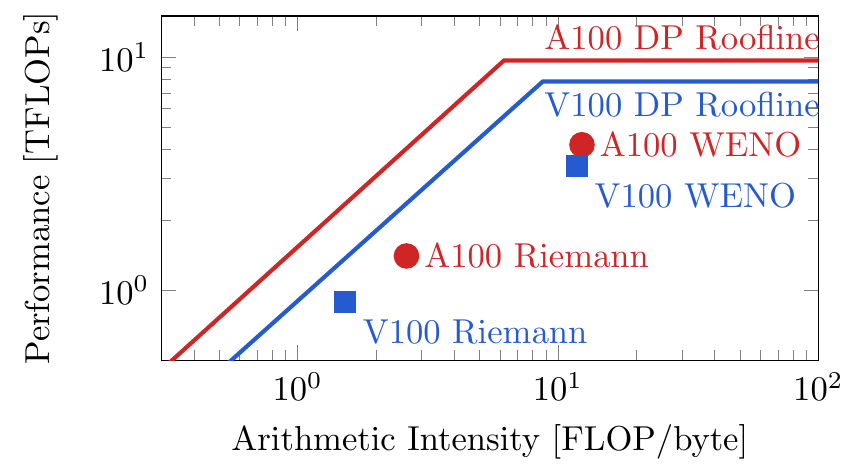}
    \caption{Roofline analysis for the two most expensive kernels, WENO and Riemann solvers. 
    The kernels and rooflines are associated with double precision (DP) computation and fused multiply-add (FMA) operations.}
    \label{fig:roofline}
\end{figure}

A roofline analysis of the two most expensive kernels, WENO reconstruction and Riemann solvers, on NVIDIA A100 and V100 GPUs and double precision rooflines using NVIDIA Nsight Compute~\citep{nsightc} are shown in \cref{fig:roofline}. 
A high arithmetic intensity of $10$~FLOP/byte and over $40$\% of the peak double precision FLOPs are observed for the WENO kernel on both GPUs. 
The kernel achieves a bandwidth of \SI{305}{\giga\byte\per\second} on the V100 GPU, with the peak bandwidth being \SI{900}{\giga\byte\per\second}. 
Kernel optimizations and high memory reuse are largely responsible for efficiently utilizing compute resources. 
In contrast, the Riemann solvers have a lower arithmetic intensity ($2$~FLOP/byte) and only achieve about $10$\% of the peak double precision FLOPs on both GPUs. 
This can be attributed to low memory reuse and an abundance of conditional statements in this kernel. 
However, efforts to alleviate performance issues by employing single-precision arithmetic in the Riemann solvers kernel are currently being pursued. 

\subsection{Architecture comparisons}

The portability of our acceleration strategies is verified by testing performance on various hardware architectures. 
We tested performance on several CPUs: Ampere Altra Q80-30 (located on OLCF Wombat), Intel Xeon Gold Cascade Lake (SKU 6248m, PSC Bridges2), and IBM Power9 (OLCF Summit). 
NVHPC and GCC11 compilers were tested with \texttt{-fast} and \texttt{-Ofast} compiler optimization flags, respectively. 
GPU performance was analyzed for the NVIDIA V100 (OLCF Summit) and A100 (OLCF Wombat) using the NVHPC~22.1 compiler with the \texttt{-Ofast} flag.
All computations are double precision.
A 3D two-component problem with $16$M grid points on 2 MPI ranks was used for testing. 
The wall time was averaged over $10$ time steps. 

\begin{table}
    \centering
    \renewcommand{\arraystretch}{1.15}
    \small
    \begin{tabular}{c c c c }
        \hline\hline
         & \# Cores &  Time [s] & Slowdown  \\
        \hline
         A100 & --- &  0.28 & Ref. \\
         V100  & --- &  0.50 & 1.7 \\
         Xeon  & 40  & 2.1\phantom{0} & 7.3 \\
         Ampere & 40 & 2.7\phantom{0} & 9.2 \\
         Power9 & 42 & 3.5\phantom{0} & 12\phantom{.0} \\
         \hline\hline
    \end{tabular}
    \vspace{0.2cm}
    \caption{Comparison of wall times per time step on various architectures.
    The Intel Xeon Gold chips are the Cascade Lake architecture, and Ampere indicates the Ampere Altra Q80-30 chip.
    The A/V100 GPU simulations use the NVHPC~v22.1 compiler, and the CPUs use GNU~v11.1 (the fastest among tested compilers).}
    \label{tab:MFC_Run}
\end{table}

\Cref{tab:MFC_Run} shows average wall times and relative performance metrics for the different hardware.
The ``Time'' column has little absolute meaning, with the relative performance being the most meaningful (also shown in the last column).
In~\cref{tab:MFC_Run}, the performance of a single GPU is compared to that of the CPU chip, with the CPU wall times normalized by the number of CPU cores per chip. 

The results show that the A100 GPU is $1.72$-times faster than the V100 on OLCF Summit, faster than even the peak double-precision performance would anticipate between the two cards (a factor of $1.24$).
This can be attributed to higher memory bandwidth ($1.7$-times) and faster GPU interconnects (2-times) on an A100 over a V100.
A single A100 also gives a $7.3$-times speedup over the Intel Xeon Cascade Lake, the fastest CPU chip.
Increasing the number of GPU ranks has a negligible effect ($\approx 15$\%) on performance due to the relatively small contribution of the MPI subroutine~\cref{s:kern}. 
The GNU11 compilers give shorter wall times than the NVHPC compilers on all CPU architectures. 
The Ampere Altra CPUs are $1.4$-times faster when compared to the POWER9s and $1.2$-times slower than the Intel Xeons. 
The Arm-based CPUs (Ampere Altra) are more energy- and cost-efficient than their x86 counterparts (Intel Xeon) under their reduced instruction set (RISC). 
While performance on x86 CPUs is superior, future trends indicate a growing popularity for Arm chips in high-performance computing applications~\citep{yokoyama2019survey}. 
The performance for different Arm chips, including the TX2 and A64FX, and their compilers are discussed for various applications, including MFC, in \citet{elwasif23}.

\begin{figure}[h]
    \centering
    \includegraphics[scale=1]{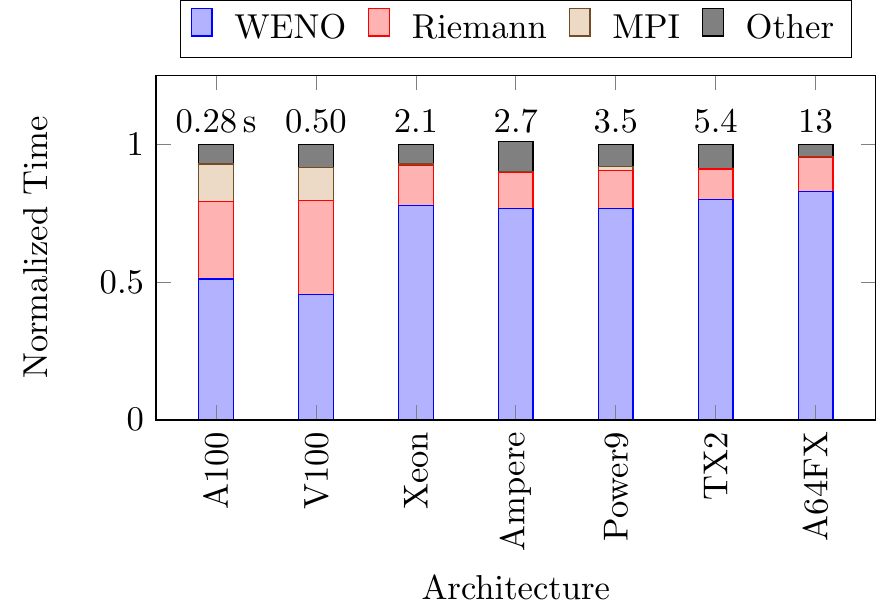}
    \caption{
        Cost breakdown of different MFC subroutines on various architectures for a 3D $2$-component problem with $16$M grid points.
        Cases V100 and A100 have all compute kernels on the respective GPUs, so the associated CPU architecture is not meaningful.
        Numbers above the bars indicate the absolute wall time in seconds, also shown in \cref{tab:MFC_Run}.
    }
    \label{fig:NormTime}
\end{figure}

\Cref{fig:NormTime} shows a timestep normalized breakdown of the duration of the most expensive MFC routines.
The two left columns indicate kernel times on GPUs; the rest are CPU-only.
All computation is offloaded to accelerators, with CPUs being used only for I/O operations and halo exchanges on systems with no support for GPU-aware MPI. 
MPI communication constitutes a meaningful proportion of the total time on GPUs while being negligible on CPUs. 
This can be attributed to much larger speedups in computation on GPUs over CPUs as compared to MPI communication. 
The relative proportion of various routines remains nearly constant across CPU and GPU architectures. 
In addition to the architectures of \cref{tab:MFC_Run}, \cref{fig:NormTime} includes results for older ARM architectures like the ThunderX2 (TX2) and Fujitsu A64FX. 
These processors are markedly slower than modern x86 or ARM chips, though the profiles are qualitatively similar to the other CPUs.

\section{Example simulations}\label{s:exmp}

We demonstrate the capability and flexibility of the method described via a large multiphase simulation of different test configurations. 
GPU speedups are consistent with those of \cref{s:main}.

\subsection{Cavitating bubble cloud}

\begin{figure}[ht]
    \centering
    {\setlength{\tabcolsep}{10pt}
    \begin{tabular}{c c}
        \includegraphics[width=2.95in]{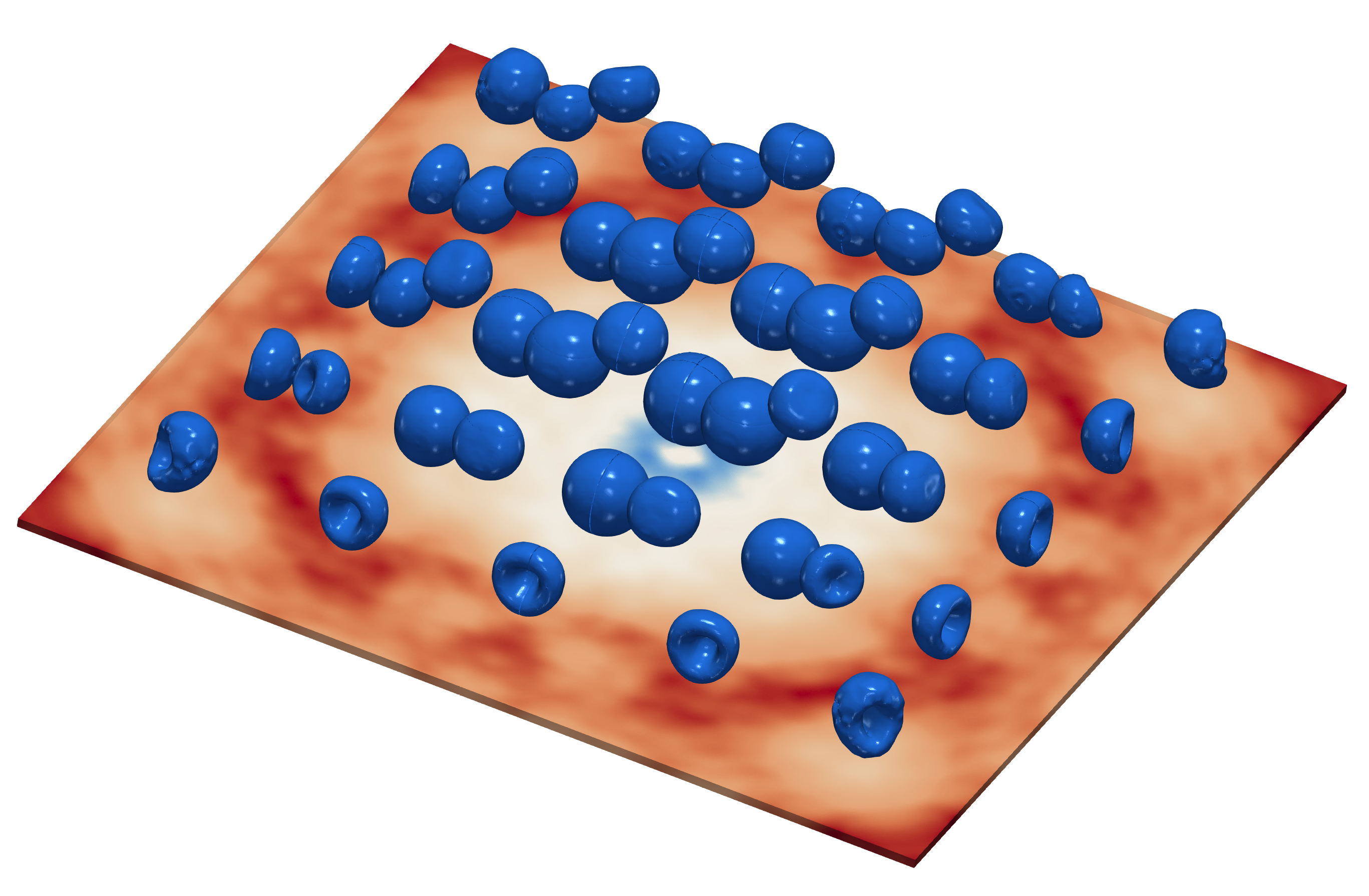} &
        \includegraphics[width=2.95in]{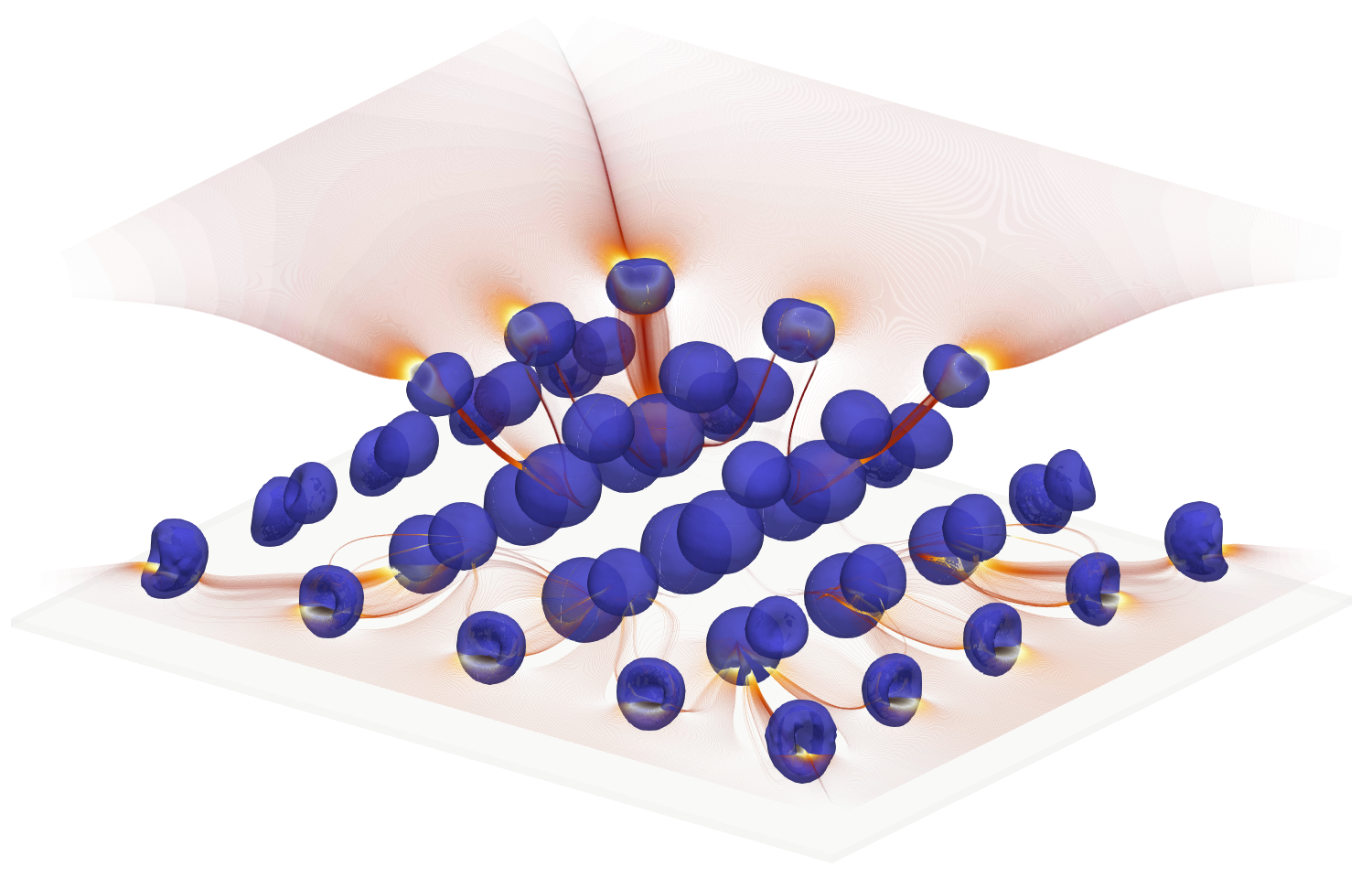} \\
        \small (a) Wall pressure & \small (b) Streamlines
    \end{tabular}
    }
    \caption{
        Illustrative MFC simulation of a collapsing bubble cloud near a wall. 
        Results show the bubbles as contours of $\alpha_1 = 0.5$ and pressures and streamlines as labeled.
        In~(a), red colors are larger pressures, and blue colors are lower.
    }
    \label{fig:mfc_cavbub}
\end{figure}

The first example case simulates the collapse of $50$ air bubbles in water at ambient pressure (\qty{1}{\atmosphere}) near a wall when subject to a \qty{10}{\atmosphere} pressure wave.
The density of the fluids follows from those at standard temperature and pressure.
The stiffened-gas equation of state represents both fluids~\citep{le2016noble}.
This example simulation uses $216$M grid points and $60$ grid points across each initial bubble diameter. 
We simulate for $3 \times 10^5$ time steps, corresponding to a dimensionless time $\tau \equiv t c / R_0 = 42$ for air sound speed $c$ and equilibrium bubble radii $R_0$.

\Cref{fig:mfc_cavbub} shows the wall pressure contours and streamlines of the collapsing bubble cloud and the isocontours of air volume fraction.  
Here, we use $216$~GPUs ($36$~nodes) on Summit, or $1$M points per GPU, which amounts to $3$~hours of wall-time for this simulation.
Using POWER9 CPUs on the same compute nodes would require more than 2 days for the same problem. 
We save the simulation state to disk every 1000~time steps.
Each data export requires about ten steps worth of wall time, so this cost has a negligible impact on performance.

\subsection{Shock--bubble-cloud--stone interaction}

The approach of this manuscript can also simulate the shock-induced collapse of air bubbles near a model kidney stone.
This, in part, demonstrates the efficiency and capabilities of the method.
The configuration of interest is similar in spirit to shock- and burst-wave lithotripsy~\citep{tanguayComputationBubblyCavitating2004, katzInvestigationEnergyShielding2018}.

Here, we consider a dispersion of 17 air bubbles initially near a model stone and submerged in water.
The impinging shock has a Mach number $M_s = 7.92$.
The BegoStone material represents the stone~\citep{liu2002begostone}, a kidney stone phantom in lithotripsy research trials~\citep{zwaschka2018combined}.
BegoStone has a density of $\rho = \SI{1995}{\kilogram\per\metre^3}$ and longitudinal and transverse wavespeeds $c_L= \SI{4159}{\meter\per\second}$ and $c_T = \SI{2319}{\meter\per\second}$.

The hypoelastic model described in \cref{ss:hyp} represents elastic stresses in the stone and carries out this simulation. 
The simulation domain is $2.67D$ in the mean flow direction and $1D$ in the transverse and spanwise directions, where $D$ is the stone's diameter. 
A structured $1600 \times 600\times 600$ Cartesian grid (576M grid points) discretizes the domain.
This simulation was conducted on 576~GPUs (96~nodes) on OLCF Summit for $25 \times 10^3$ time steps, corresponding to a wall-time of 30~minutes.

\begin{figure}[t]
    \centering
    \includegraphics[width=0.6\textwidth]{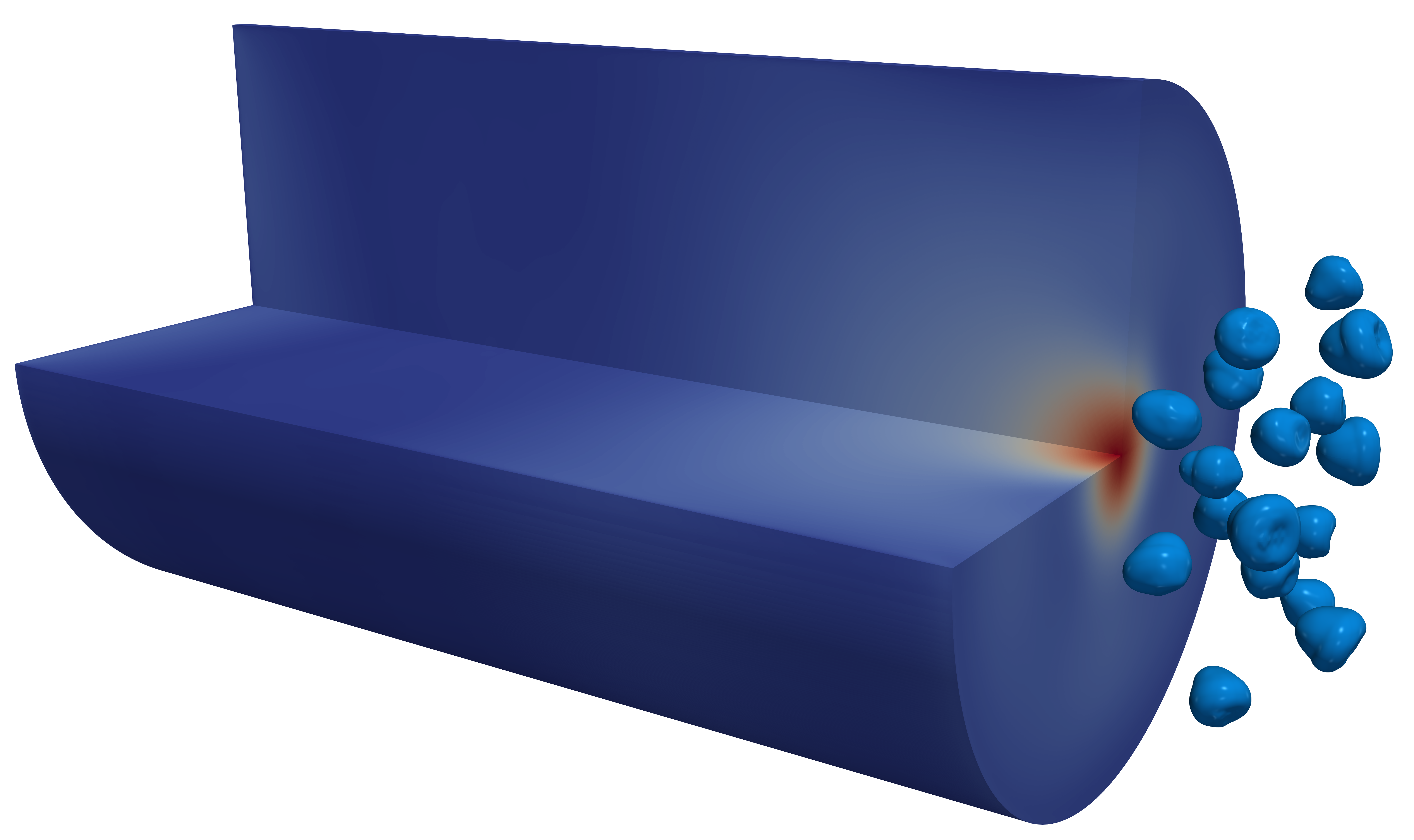}
    \caption{
        Illustrative simulation of a kidney stone near a collapsing bubble cloud. 
        Reds indicate higher stresses, and blues indicate lower stresses. 
        The bubble and stone isosurfaces are shown for volume fraction $\alpha = 0.5$.
    }
    \label{fig:mfc_hyp}
\end{figure}

\Cref{fig:mfc_hyp} shows maximum principal stresses (defined in the usual way) in the stone at $\tau \equiv t c /R_0 =  2.03$ for air sound speed $c$ and initial bubble radius $R_0$.
These stresses follow from the shock, and the later bubble collapses.
As expected, we observe larger stresses near the center of the bubble cloud and stone cross-section.

\subsection{Atomizing droplet}

The third example simulation shows the atomization of a 3D water droplet in air impinged by a Mach $1.46$ shock wave. 
The domain is 20$D$ in the mean flow direction and 10$D$ in the transverse and spanwise directions, where $D$ is the initial droplet diameter.
A $2000 \times 1000 \times 1000$ Cartesian grid (2B grid points) discretizes the domain. 
Stretching locally refines and stretches the grid near the droplet as
\begin{equation}
\begin{aligned}
    x_{\text{stretch}} = x + \frac{x}{a_x} {\bigg[} &\log\left(\cosh\left(\frac{a_x(x - x_a)}{L}\right)\right) + \\
    &\log\left(\cosh\left(\frac{a_x(x - x_b)}{L}\right)\right) - 2\log\left(\cosh\left(\frac{a_x(x_b-x_a)}{2L}\right)\right) {\bigg]}
\end{aligned}
\end{equation}
where $a_x$ is the stretch magnitude, $L$ is the domain length, and $x_a$ and $x_b$ control where stretching occurs.
The droplet is centered at the origin, with $x_a = -1.2D$ and $x_b = 1.2D$ and stretching factor $a_x = 4$ across all coordinate directions. 
This flow is simulated for $2 \times 10^5$ time steps with a dimensionless timestep $\Delta \tau = 9.6 \times 10^{-6}$, and the simulation state is saved every 1000 time steps.
The simulation is conducted on OLCF Summit using $960$ GPUs ($160$ nodes), which amounts to 4~hours of wall-time.

\begin{figure}[H]
    \centering
    \includegraphics[width=0.7\textwidth]{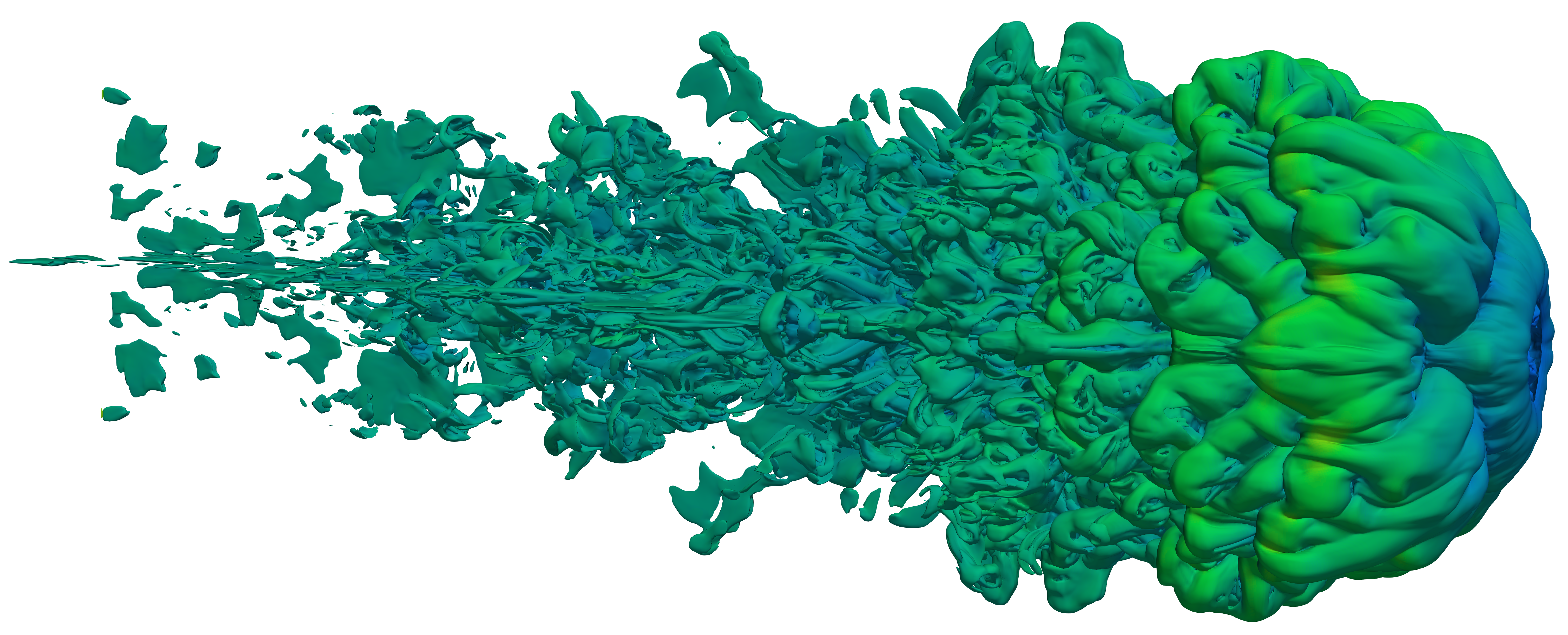}
    \caption{Vorticity isocontours $\lVert \boldsymbol\omega \rVert = 10^5$ around a shedding water droplet.}
    \label{fig:mfc_drop}
\end{figure}

\Cref{fig:mfc_drop} shows the vorticity ($\boldsymbol{\omega}$) magnitude isosurface $\lVert \boldsymbol\omega \rVert = 10^5$ at dimensionless time $\tau = 1.43$, where $\tau \equiv t u_g / D \sqrt{\rho_g/\rho_l}$, for shock velocity $u_g$ and post-shock gas and liquid densities $\rho_g$ and $\rho_l$.
The shock wave leads to two counter-rotating vorticity streams in the wake of the droplet, forming a recirculation region as shown in \cref{fig:mfc_drop}. 
This simulation agrees with previous studies of droplet aerobreakup~\citep{mengNumericalSimulationAerobreakup2018}, though it requires shorter wall times and makes judicious use of the latest accelerators.

\section{Conclusions}\label{s:conclusions}

We present a GPU-accelerated multiphase compressible flow solver, MFC, capable of simulating various physical phenomena. 
Optimization techniques are used to facilitate efficient memory use in computationally expensive kernels. 
This enables higher arithmetic intensity over typical CFD algorithms for compressible flow.
Further performance improvements are observed through metaprogramming techniques via Fypp. 
Portability of the implementation is ensured through directive-based offloading via OpenACC, and performance is tested on NVIDIA GPUs as well as Intel Xeon, IBM POWER9, and ARM CPUs. 
A 40-times speedup is observed on a single Summit node using NVIDIA V100 GPUs over POWER9 CPUs. 

Multi-GPU performance is examined by conducting weak and strong scaling tests. 
Near ideal weak scaling (with $3\%$) is observed for up to 13824 GPUs on Summit. 
CUDA-aware MPI coupled with the GPUDirect RDMA further reduces communication overhead on GPUs, reducing simulation time for all cases tested.
Level-3 MPI parallel I/O ensures that the cost of data dumps is negligible for large simulations. 
Scaling and speedup results show significant improvement over previous multiphase implementations~\citep{crialesi2023flutas} and are comparable to previous single-phase implementations~\citep{romero2020zefr, bernardini2021streams}. 

The ability to conduct large multiphase simulations was tested by conducting large multi-GPU simulations across many compute nodes. 
The example simulations performed $\mathcal{O}(10^5)$ time steps within a few hours on NVIDIA GPUs instead of a few days on current x86 and ARM multicore CPU processors.
This strategy is poised to enable the efficient use of existing and upcoming exascale systems like OLCF Frontier and LLNL El Capitan for state-of-the-art compressible multiphase flow simulations.

\section*{Acknowledgments}

We acknowledge useful discussions of this work from Brent Leback and Mat Colgrove (NVIDIA), St\'{e}phan Ethier (Princeton), Nicholson Koukpaizan (Oak Ridge National Lab), Pedro Costa (TU~Delft), and Luca Brandt (KTH, Sweeden).
SHB acknowledges the support of this work via the US Office of Naval Research under grant number N00014-22-1-2519 (PM Julie Young), hardware gifts from the NVIDIA Corporation, and use of OLCF Summit and Wombat under allocation CFD154.
TC acknowledges support via the US Office of Naval Research under grant number N00014-22-1-2518 (PM Julie Young).
This work used Bridges2 at the Pittsburgh Supercomputing Center through allocation PHY210084 from the Advanced Cyberinfrastructure Coordination Ecosystem: Services \& Support (ACCESS) program, which is supported by National Science Foundation grants \#2138259, \#2138286, \#2138307, \#2137603, and \#2138296.

\bibliographystyle{model1-num-names}
\bibliography{manuscript.bib}
\end{document}